\documentstyle[12pt,twoside,emlines,cmp209]{article}
\title[Multifractal Spectra]{Multifractal Dimension Spectra in
Polymer Physics}
\author[Christian von Ferber, Yurij Holovatch]{Christian von
Ferber\refaddr{label1,label2}, Yurij Holovatch\refaddr{label3}}
\addresses{
\addr{label1} Institut f\"ur Theoretische Physik, Heinrich
Heine Universit\"at D\"usseldorf, D-40225 D\"usseldorf, Germany
\addr{label2}School
of Physics and Astronomy, Tel Aviv University, IL-69978 Tel-Aviv,
Israel
\addr{label3} Institute for Condensed Matter Physics, Ukrainian Acad.
Sci., 1 Svientsitskii Str., UA-290011 Lviv, Ukraine
}
\begin{document}

\maketitle

\begin{abstract}
We study the multifractal properties of diffusion in the presence
of an absorbing polymer and report the numerical values of the
multifractal dimension spectra for the case of an absorbing
self avoiding walk or  random walk.
\keywords Multifractals, Spectral Function, H\"older Exponent.
\pacs 6460.Ak, 61.41.+e, 64.60.Fr, 11.10.Gh
\end{abstract}

\section{Introduction}\label{I}
The notion of non-integer dimension although developed by
mathematicians in late 19th - early 20th centuries has found its wide
application in physics only in the last few decades. This application
was mainly initiated by the pioneering work of B. Mandelbrot who attracted
attention to the fact that ``fractal'' geometry provides an
appropriate framework for the description of a whole range of complex
structures that are formed from smaller subunits \cite{Mandelbrot83}. Whereas
such structures are characterized by appropriate fractal dimensions,
their growth and spatial correlations
\cite{growth} are described by a (non-trivial)
spectrum of multifractal (MF) dimensions \cite{Hentschel83,Halsey86}.

In this paper, we study the properties of diffusion phenomena in the
presence of an absorbing polymer. This provides another example of a
multifractal phenomenon in condensed matter physics. To this end we use
the model proposed by Cates and Witten \cite{Cates87} and
derive the MF spectrum in the frames of a field theoretical formalism
using the renormalization group \cite{Bogoliubov59,Zinn89} method.
The MF spectrum is related to the spectrum of exponents governing
scaling properties of co-polymer stars \cite{FerHol97}. We calculate
this spectrum to the third order of perturbation theory and report
the numerical values of the  quantities that characterize the MF behaviour.

\section{The Model and the Multifractal Measure
}\label{II}
\begin{figure} [htbp]
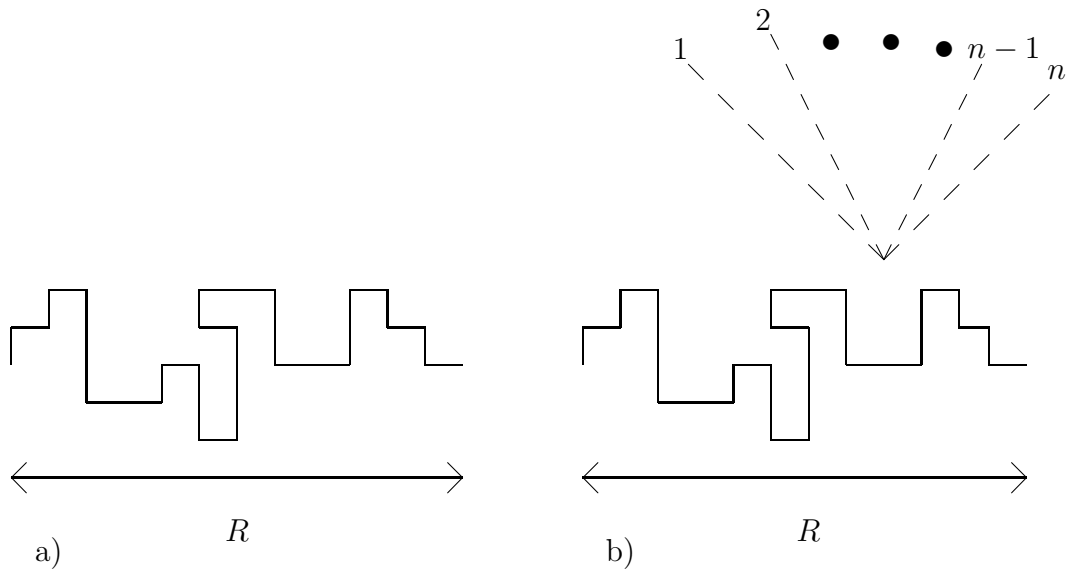

\begin{center}
\input fig1.pic
\end{center}
\caption{ \label{fig1}
a) A SAW of $N$ steps each
of size $a$ in $d$-dimensional space.
In the asymptotic limit $N \rightarrow \infty$
the mean square distance between the
extremities is related to the `mass' $N$ by
 $N \simeq (R/a)^{1/\nu}$ with $\nu(d=2)=3/4$,
$\nu(d=3)=0.588$. This defines the fractal dimension of a SAW as $D=4/3$,
$D=1.70$ in two and three space dimensions.
b) A star of $n$ RW (trajectories of diffusing particles) that reach the
same point close to the
absorbing polymer. The averaged $n$th moment of concentration
of diffusing particles $\rho$ scales as
$<\rho^n> \sim (R/a)^{-\lambda_n}$. The set of scaling dimensions
$\lambda_n$ depends on $n$ in a non-trivial way, leading to the
non-trivial behaviour of the flux moments of diffusing particles
on the surface of fractal (polymer)  absorber.
}
\end{figure}

Long flexible polymer chains \cite{note1}
provide a classical example of fractal
objects (see fig. \ref{fig1}a).
Of particular interest is the manner
in which such a chain absorbs diffusing particles. The flux of
particles on the surface of the absorbing chain is defined by
the density of incoming particles close to the absorber
(see fig. \ref{fig1}b).
Taken $\rho(r)$ to
be the probability density of incoming particles the problem is
to solve the steady-state diffusion equation (Laplace
equation)
\begin{equation}\label{1}
\nabla^2 \rho(r) = 0
\end{equation}
with boundary conditions for a given absorbing polymer
\begin{eqnarray}\nonumber
\rho(r)&=&0 \hspace{5em} \mbox{on the surface of the absorber}
\\ \label{2}
\rho(r)&=&\rho_{\infty}=const \hspace{5em} \mbox{ for } \quad |r| \to
\infty .
\end{eqnarray}
One motivation for the study of this problem is to gain insight to the
description of diffusion limited aggregation (DLA) \cite{Witten81}.
The last phenomenon is much more complicated because of the fact
that for DLA the boundary conditions are given on the surface of
growing aggregate itself. The process described by equations (\ref{1}),
(\ref{2}) may be rather considered as diffusion limited catalysis,
when particles of one type interact with a {\it prescribed} fractal
(a catalysing polymer) and transform into the other type \cite{Oshanin}.

Let the absorbing polymer of size $R$ be chosen from the well defined
ensemble of SAW. Then we are interested in the moments of the field
$\rho(r)$ close to the surface of the absorber (see fig. \ref{fig1}b).
For distances $R>>r>a$ ($a$ being a cut off) the averaged moments
$<\rho(r)^n>$ are expected to scale as \cite{Cates87}
\begin{equation} \label{2a}
<\rho(r)^n> \sim (R/r)^{-\lambda(n)},
\end{equation}
where $<\dots>$ denotes the average over the ensemble of polymers.
Let us take that the flux $\phi(x)$ onto any randomly chosen
point $x$ of the absorber is proportional to the field $\rho(a)$ at
a point that is as close as a cut off length $a$ from the absorber.
Then, for the averaged moments of the flux one finds:
\begin{equation} \label{3}
<\phi^n> \sim (R/a)^{-\lambda(n)}.
\end{equation}
Following \cite{Cates87}  we associate with the flux of the Laplacian field
a ``harmonic measure'' $\mu(x)$ defined on the set of absorbing sites $x$ :
\begin{equation} \label{4}
\mu(x) = \phi(x) / \sum \phi(x),
\end{equation}
where the sum in (\ref{4}) spans all sites $x$ of the
absorber. In particular, the mass $M$ of the absorber is given by
$\sum_x1=M$ and (\ref{4}) may be rewritten as
\begin{equation} \label{5}
\mu(x) = \phi(x) / (M<\phi>),
\end{equation}
where $<\dots>$ denotes the site average $<\dots>=\sum(\dots)/M$
\cite{note2}.

It is standard \cite{Hentschel83} to describe the properties of the harmonic
measure by a {\it set of exponents} $D(n)$ given by:
\begin{equation} \label{6}
\sum\mu(x)^n = (R/a)^{(1-n)D(n)}.
\end{equation}
Note, that for $n=0$ $D(0)$ is the fractal dimension of the
absorber. Multifractal scaling is found when the spectrum $D(n)$
is non-trivial: $D(n)\neq D(0)$, or correspondingly (c.f. eq.
(\ref{3})) $\lambda (n)\neq n\lambda (1)$.  Note as well that from
(\ref{3}), (\ref{6}) the exponents
$\lambda(n)$ and $D(n)$ are related:
\begin{equation} \label{7}
n\lambda(1)-\lambda(n) = (1-n)[D(n)-D(0)].
\end{equation}
In many cases the description of a multifractal measure is performed in
terms of the {\it H\"older exponent} $\alpha$  and the {\it spectral
function} $f(\alpha)$ that are defined by a Legendre transform
of $D(n)$ \cite{Halsey86}:
\begin{eqnarray}
\alpha(n) & = & \frac{d}{dn}[(n-1)D(n)],
\label{8} \\
f(\alpha) & = & n \alpha(n) + (1-n)D(n). \label{9}
\end{eqnarray}
The function $f(\alpha)$ differs for
different fractal measures but it possesses certain `universal'
properties.
In particular, its maximum $\max_{\alpha} f(\alpha)$ gives the fractal
dimension of the absorber and $f$ is convex $f''(\alpha)< 0$.

\section{The Multifractal Spectrum
Exponents and the Spectral Function}\label{III}
The central idea that allows us to calculate the properties of the solutions
of equation (\ref{1}) with the boundary conditions (\ref{2}) is that
a path integral representation both for the field $\rho(r)$
and for the absorber (being a RW or a SAW) is possible \cite{Cates87,Oshanin}.
In terms of the path integral solution of the Laplace equation one
finds that $\rho(r)$ at point $r$ near absorber is proportional to the
number of RW that end at point $r$ and avoid the absorber.
The $n$th power of this field is proportional to the $n$th power of
the above mentioned number, i.e. it is defined by a partition function
of a star polymer with $n$ arms \cite{stars}
(the latter is shown by dashed lines in
fig. \ref{fig1}b). Furthermore, introducing the mutual avoidance
conditions between the ``$n$-arm star'' and the 2-arm polymer (representing
the absorber) one has to calculate the partition function of a co-polymer
star consisting of chains of two different species that avoid each other.
These correspond to the
trajectories of diffusing particles (being RW) and the absorbing polymer
(which for the purpose of present study is chosen as RW or SAW).
Making use of the previously developed \cite{FerHol97}
theory of copolymer stars and networks and mapping the model of
co-polymer stars to the appropriate field theory
\cite{Schaefer91,FerHol96b}
one may relate \cite{FerHoldubna} the MF spectrum exponents (\ref{3})
to the exponents that define the  scaling properties of co-polymer
stars. In particular, for a co-polymer star consisting of $n_1$ chains
of species $1$ and $n_2$ chains of species $2$ the number of
configurations $Z_*$ scales like \cite{FerHol97}:
\begin{equation} \label{10}
Z_* \sim (R/a)^{\eta_{n_1n_2}-n_1\eta_{20} -n_2\eta_{02}}.
\end{equation}
Here, the $\eta_{n_1n_2}$ represent the co-polymer star
exponents. The latter have been calculated using a field theoretical
renormalization group approach
\cite{Bogoliubov59,Zinn89}
and are known in the third order of
perturbation theory in $\varepsilon=4-d$ and pseudo-$\varepsilon$
\cite{note3}
expansions \cite{FerHol97,FerHol98a}. By means of a
short-chain expansion \cite{Ferber98} the set of exponents
$\eta_{n_1n_2}$ can be related to the exponents $\lambda(n)$
(\ref{3}), that govern the scaling of
the $n$th moment of the
flux onto an absorbing linear chain. Considering the absorber to be
either a RW or a SAW we define the exponents:
\begin{eqnarray} \label{11}
\lambda_{RW}(n) \def \lambda_{2,n}^G=-\eta^{G}_{2n}, \\
\label{12}
\lambda_{SAW}(n) \def
\lambda_{2,n}^U=-\eta^{U}_{2n}+\eta_{20},
\end{eqnarray}
here, we keep notations of
\cite{FerHol97,FerHol98a}
for the expressions for
exponents $\eta$ defined in different fixed points of the
renormalization group transformation. Based on these expressions we
get:
\begin{eqnarray}
\lambda_{RW} (\varepsilon) &=&
\varepsilon\,n-{\frac {n\left (n-1\right ){\varepsilon}^{2}}{4}}+{\frac
{n\left (n-1\right )\left (-1+n+3\,\zeta (3)\right ){\varepsilon}^{3}
}{8}} ,
\label{13}
\end{eqnarray}
\begin{eqnarray} \nonumber
\lambda_{SAW} (\varepsilon) &=&
\frac {3\varepsilon n}{4} +
\frac{n(7n-18)\varepsilon^2}{128}
+ \\ &&
\frac{n(-149-42n+108n^2-276\zeta(3)+540\zeta(3)n)\varepsilon^3}{2048}.
\label{14} \end{eqnarray}
Here, $\varepsilon=4-d$,  $d$ is the space dimension, and $\zeta(3)
\simeq 1.202$ is the Riemann zeta function.

\begin{table} [tb]
\caption {
\label{tab1}
Exponents $\lambda_{RW}(n)$ and $\lambda_{SAW}(n)$ obtained in
$\varepsilon$ and in pseudo-$\varepsilon$ expansion ($\lambda(\tau)$)
techniques.
}
\begin{centering}
\begin{tabular}{llllll}
\\
\\
\hline
\\
&$n$
& $\lambda_{RW}(\varepsilon)$ & $\lambda_{RW}(\tau)$
& $\lambda_{SAW}(\varepsilon)$ & $\lambda_{SAW}(\tau)$  \\
&1 &  0.99 &  0.99 & 0.71 & 0.71 \\
&2 &  1.77 &  1.81 & 1.31 & 1.33 \\
&3 &  2.45 &  2.53 & 1.86 & 1.92 \\
&4 &  3.01 &  3.17 & 2.34 & 2.44 \\
&5 &  3.51 &  3.75 & 2.78 & 2.94 \\
&6 &  3.95 &  4.28 & 3.19 & 3.41 \\
\\
\hline
\end{tabular} \\
\end{centering}
\end{table}

Using the perturbation expansions for the $\lambda$ exponents
and the relations for $\lambda(n)$
and the spectral function some algebra leads to the corresponding
expansions for $\alpha_n$ and $f(\alpha_n)$:
\begin{eqnarray}
\alpha_{RW} (\varepsilon) &=&
2 + \frac{(-2n+1)\varepsilon^2}{4} +
\frac{(-4n-3\zeta(3)+3n^2+6\zeta(3)n+1)\varepsilon^3}{8},
\label{15}
\end{eqnarray}
\begin{eqnarray}
f_{RW} (\varepsilon) &=&
2  - \frac {\varepsilon^2 n^2}{4} +
\frac{(2n^3+3\zeta(3)n^2-2n^2)\varepsilon^3}{8} ,
\label{16}
\end{eqnarray}
\begin{eqnarray} \nonumber
\alpha_{SAW} (\varepsilon) &=&
2 - \frac{\varepsilon}{4} +
\frac{(7-36n)\varepsilon^2}{128}+
\\ &&
\frac{(-149-276\zeta(3)-84n+324n^2+1080\zeta(3)n)\varepsilon^3}{2048},
\label{17}
\end{eqnarray}
\begin{eqnarray}    \nonumber
f_{SAW} (\varepsilon) &=&
2 - \frac{\varepsilon}{4} -
\frac{(18n^2+11)\varepsilon^2}{128}+
\\ &&
\frac{(216n^3+540\zeta(3)n^2-42n^2-83+264\zeta(3))\varepsilon^3}{2048}.
\label{18}
\end{eqnarray}
The exponents $\lambda(n)$ as well as the H\"older exponents $\alpha$
and the spectral functions $f(\alpha)$ in terms of the
pseudo-$\varepsilon$ expansion \cite{note3} are given elsewhere
\cite{FerHoldubna,FerHol98b}.

As is well known, the series of type (\ref{13})-(\ref{18}),
as they occur in field theory appear
to be of asymptotic nature with zero radius of convergence. However,
knowledge of the asymptotic behavior of the series as
derived from the renormalization group theory allows us to evaluate these
asymptotic series (see e.g.\cite{Zinn89}). To this end
several procedures are available depending on the additional
information known for the series to be resummed.  We use the Borel
summation technique improved by a conformal mapping procedure that
has served as a powerful tool in field theoretic calculations (see
\cite{LeGuillou80} for example).

\begin{figure} [htbp]
\begin{center}
(Fig.2a)\input fig2a.pic\\[5mm]
(Fig.2b)\input fig2b.pic
\end{center}
\caption{ \label{fig2}
The spectral function $f(\alpha)$ for absorption on (a) $RW$ and (b) $SAW$.
Solid curves: 1 - [2/1] Pad\'e approximant for $\varepsilon^3$ results,
2- [2/1] Pad\'e approximant for pseudo-$\varepsilon^3$ results;
dashed curves: 3 - $\varepsilon^2$ results without resummation,
4 - pseudo-$\varepsilon^2$ results without resummation;
stars - resummed $\varepsilon^3$ results;
boxes - resummed pseudo-$\varepsilon^3$ results.
}
\end{figure}
Table \ref{tab1} contains the results for the exponents
$\lambda_{RW}(n)$ and $\lambda_{SAW}(n)$ obtained in $\varepsilon$ and
in pseudo-$\varepsilon$ expansion techniques at $d=3$ from the
corresponding values of exponents $\eta_{f_1,f_2}$  \cite{FerHol97}
with the application of this resummation procedure.
\section{Conclusions}\label{IV}

We have calculated the spectral function that describes the
scaling of the moments of measure
defined by diffusion near an absorbing polymer.
These moments were calculated  as averages over
all configurations of the absorber instead of performing a site average.
Thus the interpretation of $f(\alpha)$ itself does not
directly correspond to the standard
picture, where $f(\alpha)$ is interpreted as the fractal dimension of a
particular subset of the measure \cite{Halsey86}.
Nevertheless we derive $f(\alpha)$ in the same way from the
spectrum of scaling exponents of the moments of a measure. As we will
see below in the general properties of $f(\alpha)$  also hold
for the present definition.

Our numerical results for the spectral function for space dimension $d=3$
are presented in Figs. \ref{fig2}a,\ref{fig2}b.
They were obtained from the series for $\alpha_n$ and $f(\alpha_n)$ as
functions of $n$. We show the results of the
resummation procedure described above applied to the
series in $\varepsilon$- and
pseudo-$\varepsilon$- expansion.
For comparison we also show the curve
for direct summation of the $\varepsilon$ and pseudo-$\varepsilon$
series to the 2nd order (here we recover the second-order results of
\cite{Cates87}). In addition we have performed an analytical
continuation of our series in the form of a [2/1] Pad\'e approximant for
the third order series.  It is obvious that the direct summation of the
last series fails to converge and gives reasonable values for
$\alpha_n,f(\alpha_n)$ only for small values of $n$, i.e. near the
maximum of $f(\alpha)$ at $n=0$.  The symmetry of the Pad\'e
approximant holds only in the region shown and may be an artifact of
the method. On the left wing, where it coincides with the resummed
results the Pad\'e approximant provides a continuation that is
compatible with the estimation for the minimal $\alpha$ value
$\alpha_{\min}= d-2$. The Pad\'e result is
$\alpha_{\min}(\varepsilon)=1.333$, $\alpha_{\min}(\tau)=1.017$
for the RW absorber and
$\alpha_{\min}(\varepsilon)=1.250$, $\alpha_{\min}(\tau)=1.013$
for the SAW absorber, which is calculated here only from 3rd order
perturbation theory.

As can be extrapolated from the Pad\'e approximant and as was shown also
on the basis approximations for higher moments $n$ \cite{Cates87}, $f(\alpha)$
as it is defined here, will become negative near $\alpha_{\min}$ and
$\alpha_{\max}$. For this reason the identification of $f(\alpha)$ as the
fractal dimension of some identifiable subset is not possible here.
Also the extrapolation of the resummed data seems to indicate such a
behavior. Note, however, that the perturbative approach, even in combination
with resummation and analytical continuation is still not capable to give
reliable results for high values of the expansion parameters. In
particular this method is only good near the maximum of $f(\alpha)$
\cite{note4}.

Though the
results obtained for $\alpha_n$ and $f(\alpha_n)$ for a specific value of $n$
differ in both approaches, the same curve $f(\alpha)$ is described with
better coincidence for the left wing of the curves, corresponding to
positive $n$.
The resummation techniques we apply have proven to be a powerful tool
already in the field theoretical approach to critical phenomena and have
lead to high precision values for critical exponents.
We believe that the application of these methods to calculations of MF
phenomena also allows to improve the reliability and comparability of these
results.

A generalization of our present approach to the
study of spectral functions that describe the absorption at the core
of an absorbing polymer star with a given number of arms will be published
elsewhere.

It is our pleasure to acknowledge discussions with Lothar
Sch\"afer, Bertrand Duplantier, Alexander Olemskoi. We are indebted
to Gleb Oshanin for attracting our attention to the references
\cite{Oshanin}. Part of this work was supported by DFG (SFB 237) and the
Minerva foundation.


\begin{thebibliography}{99}
\bibitem{Mandelbrot83}
Mandelbrot B.~B. The fractal geometry of nature. New York, W.H.Freeman
and Co, 1983.

\bibitem{growth}
Meakin~P. The growth of fractal aggregates and their fractal
measures. -- In: Phase transitions and critical  phenomena.
Ed. by  C.~Domb and J.~L.~Lebowitz. Vol.~12, New York,
Academic Press, 1988. p.~336--442;
Meakin~P.
Simulation of non-equilibrium growth and agregation processes. --
In: Springer Proceedings in Physics. Computer Studies in
Condensed Matter Physics. Ed.
by  D.~P.~Landau, K.~K.~Moon, H.-B.~Sch\"uttler. Vol.~33,
Berlin, Springer-Verlag, 1988. p.~55--64;
Family~F. Growth by gradients: fractal grows and pattern formation in
a Laplacian field. -- ibid., p.~65--75.

\bibitem{Hentschel83}
Hentschel H.~G.~E. , Procaccia~I.
The infinite number of generalized dimensions of fractals and
    strange attractors. // Physica~D, 1983, vol~8, p.~435-444.

\bibitem{Halsey86}
Halsey~T.~C., Jensen~M. H., Kadanoff~L.~P., Procaccia~I., and B.~I.
Shraiman~B.~I.
Fractal measures and their singularities: The characterization of
   strange sets. // Phys.~Rev.~A, 1986, vol~33, No~2, p.~1141-1151.

\bibitem{Cates87}
Cates~M.~E., Witten~T.~A.
Diffusion near absorbing fractals: Harmonic measure exponents for
   polymers. // Phys.~Rev.~A, 1987, vol~35, No~4,
p.~1809-1824.

\bibitem{Bogoliubov59}
Bogoliubov~N.~N., Shirkov~D.~V.
Introduction to the theory of quantized fields.
New York, Wiley \& Sons, 1959.

\bibitem{Zinn89}
Zinn-Justin~J.
Euclidean field theory and critical phenomena.
New York, Oxford University Press, 1989.

\bibitem{FerHol97}
von~Ferber~C., Holovatch~Yu. Copolymer Networks: Multifractal
dimension spectra in polymer field theory. // Europhys. Lett.,
1997,  vol~39, No~1, p. 31-36 ({\em preprint cond-mat/9705273});
von~Ferber~C., Holovatch~Yu. Copolymer Networks and Stars:
Scaling Exponents. // Phys.~Rev.~E, 1997, vol~56, No~6, p.~6370-6386
({\em preprint cond-mat/9705278v2}).

\bibitem{note1}
For the purpose of present study polymer chains are considered as
random walks (RW) or self-avoiding walks (SAW).

\bibitem{Witten81}
Witten~Jr.,~T.A., Sander L.M. Diffusion-limited aggregation, a
kinetic critical phenomenon. // Phys.~Rev.~Lett., 1981, vol~47,
No~19, p.~1400-1403;
Witten~Jr.,~T.A.,  Sander L.M. Diffusion-limited aggregation. //
Phys.~Rev.~A, 1983, vol~ 27, No~9, p.~5686-5697.

\bibitem{note2} In this study, the site average (\ref{5})  is
substituted by an ensemble average described by (\ref{2a}). Possible
consequences of such substitution are discussed below.

\bibitem{Oshanin} Burlatsky~S.~F., Oshanin~G.~S., Likhachev~V.~N.
Diffusion-controlled reactions involving polymer chains. //
Sov.~J.~Chem.~Phys., 1991, vol~7(7), p.~1680-1698;
Burlatsky~S.~F., Oshanin~G.~S.
Diffusion-controlled reactions with polymers. //
Phys.~Lett.~A, 1990, vol~145, No~1, p.~61-65;
Oshanin~G., Moreau~M., Burlatsky~S.
Models of chemical reactions with participation of polymers. //
Anv.~Coll.~Int.~Sci., 1994, vol~49, p.~1-46.

\bibitem{stars}
Duplantier~B.
Polymer network of fixed topology: renormalization, exact critical
exponent $\psi$ in two dimensions, and $d=4- \epsilon$. //
Phys.~Rev.~Lett., 1986, vol~57, No~8, p.~941-945;
Duplantier~B.
Statistical mechanics of polymer networks of any topology. //
J.~Stat.~Phys., 1989, vol~54, Nos~3/4, p.581-680;
Sch{\"a}fer~L., von~Ferber~C., Lehr~U., Duplantier~B.
Renormalization of polymer networks and stars. //
Nucl.~Phys.~B, 1992, vol~374, No~3, p.~473-495.

\bibitem{Schaefer91}
Sch{\"a}fer~L., Lehr~U., Kappeler~C. Higher order calculations of
the renormalization group flow for multicomponent polymer solutions.
// J.~Phys.~(Paris)~I, 1991, vol~1, p.~211-233.

\bibitem{FerHol96b}
von~Ferber~C., Holovatch~Y., Sch{\"a}fer~L.
Diffusion near an absorbing polymer. // Cond. Matt. Phys.,
1996, iss.~7, p.~15-25.

\bibitem{FerHol98a}
von~Ferber~C., Holovatch~Yu. Copolymer Networks: The spectrum of
scaling dimensions. // Physica~A, 1998, vol~249, p.~327-331.

\bibitem{FerHoldubna}
von~Ferber~C., Holovatch~Yu.,
Field-theoretic operators for multifractal moments.
-- In: Proc. of the 3rd Int. Conf.
Renormalization Group'96, Dubna, 1997, p.~123-137
({\em preprint cond-mat/9705274}).

\bibitem{note3}
Whereas the $\varepsilon$-expansion
corresponds to collecting perturbation theory terms of the same powers
of $\varepsilon=4-d$, in the pseudo-$\varepsilon$ expansion
series at each power of the pseudo-$\varepsilon$ parameter
one collects contributions from the dimension-dependent loop integrals
of the same order.

\bibitem{Ferber98}
von~Ferber~C. Operator product expansion on a fractal: The short
chain expansion for polymer networks. // Nucl.~Phys.~B, 1997, vol~490,
p.~511-542.

\bibitem{FerHol98b}
von~Ferber~C., Holovatch~Y., preprint, 1998.

\bibitem{LeGuillou80}
Le~Guillou~J.~C., Zinn-Justin~J.
Critical exponents from field theory. //
Phys.~Rev.~B, 1980, vol~21, No~9, p.~3976-3998.

\bibitem{note4}
The possible negative values of the spectral function were discussed
already in \cite{Cates87} and a physical interpretation of $f(\alpha)$
was given as a histogram of the measure $\mu$ plotted in logarithmic
variables. In this interpretation negative $f(\alpha)$ indicate that
the number of sites with a certain logarithmic measure
$\alpha\sim\ln\mu$ decreases as the size of the absorber $R$
increases. Thus for large $R$ this number can only be defined by an
ensemble average, as performed here.

\end{thebibliography}
\end{document}